\documentstyle[epslf6682,aps]{revtex}

\twocolumn

\begin{document}
\title{4/3-Law of Granular Particles\\ Flowing through a Vertical Pipe}
\author{Osamu Moriyama, Naoya Kuroiwa and Mitsugu Matsushita}
\address{Department of Physics, Chuo University, Kasuga, Bunkyo-ku, 
Tokyo 112-8551, Japan
}
\author{Hisao Hayakawa}
\address{Graduate School of Human and Environmental Studies, Kyoto University, 
Sakyo-ku, Kyoto 606-8501, Japan
} 
\date{Received June 16, 1997}
\maketitle

\abstract
Density waves of granular material (sand) flowing through a vertical pipe 
have been 
investigated.  Clear density waves emerge when the cock attached to bottom 
end of the pipe is closed.  The FFT power spectra were found to show a stable 
power-law form 
$
P(f) \sim f^{-\alpha}.
$  The value of the exponent 
was evaluated as $\alpha \cong 4/3$.  We also introduce a 
simple one-dimensional model which 
  reproduces $\alpha = 4/3$ from both simulation and theoretical analysis.
\endabstract

Recently much attention has been paid to the dynamics and statistics of 
granular materials because of their ubiquity in nature and the application 
to technology.  Unlike usual solids, liquids or gases, granular materials 
are known to show complex dynamical behaviors\cite{jaeger96}, such as 
convection\cite{taguchi}, size segregation\cite{rosato87}, bubbling
\cite{pak94}, standing waves and localized excitations under vertical 
vibration\cite{swinney95,swinney96} and a fluidized bed due to air injected 
inside a box containing granules\cite{batchelor88,sasa92}.

Pattern formation of grains flowing through a 
vertical pipe which can be regarded as a one dimensional realization of
a fluidized bed
is also a typical example of unusual features of granular motion
\cite{peng94,peng95,horikawa95,horikawa96,nakahara97}.  
Emergence of density waves ({\it e.g.} slugging) has been investigated by 
molecular dynamics (MD) and 
lattice-gas automata (LGA) simulations\cite{peng94,peng95} and by the 
experiments using sand in air 
\cite{horikawa95,horikawa96}
and 
metallic spheres in liquids\cite{nakahara97}.  Power-law form of the power 
spectrum $P(f) \sim f^{-\alpha}$, where $f$ is frequency, of density 
fluctuations was also found in both numerical simulations\cite{peng94} and 
experiments\cite{horikawa95,horikawa96}.  Although their interpretations 
on the origin of the emergence of density waves are different, estimated 
values of the exponent $\alpha$ is close to each other 
($1.3 < \alpha < 1.5$).

  Although the previous experiment\cite{horikawa95,horikawa96} reported 
$\alpha \cong 1.5$, some of their experimental procedures seem a little 
ambiguous: Since they merely plugged up the bottom hole of a pipe by half in 
order to induce density waves, the rate of air flow out of the 
bottom end of the pipe was not well controlled.  Besides, the power spectra 
they obtained were still noisy.  In this Letter we will first present 
better-controlled 
air flow out of the pipe and more accurate experimental results than the 
previous ones by increasing the number of trials.  
One of our results is the precise estimation of 
the scaling exponent of the power spectrum $P(f) \sim f^{-\alpha}$ .  The 
result is $\alpha \cong 4/3$.  We also propose a one-dimensional 
model supplemented by 
the white noise which reproduces $\alpha = 4/3$  near the 
neutral curve of the linear stability analysis of uniform states.
 In other words, we will clarify 
the origin of the power-law in density waves of granular pipe flows. 

\begin{figure}[htbp]
\begin{center}
\epsfile{file=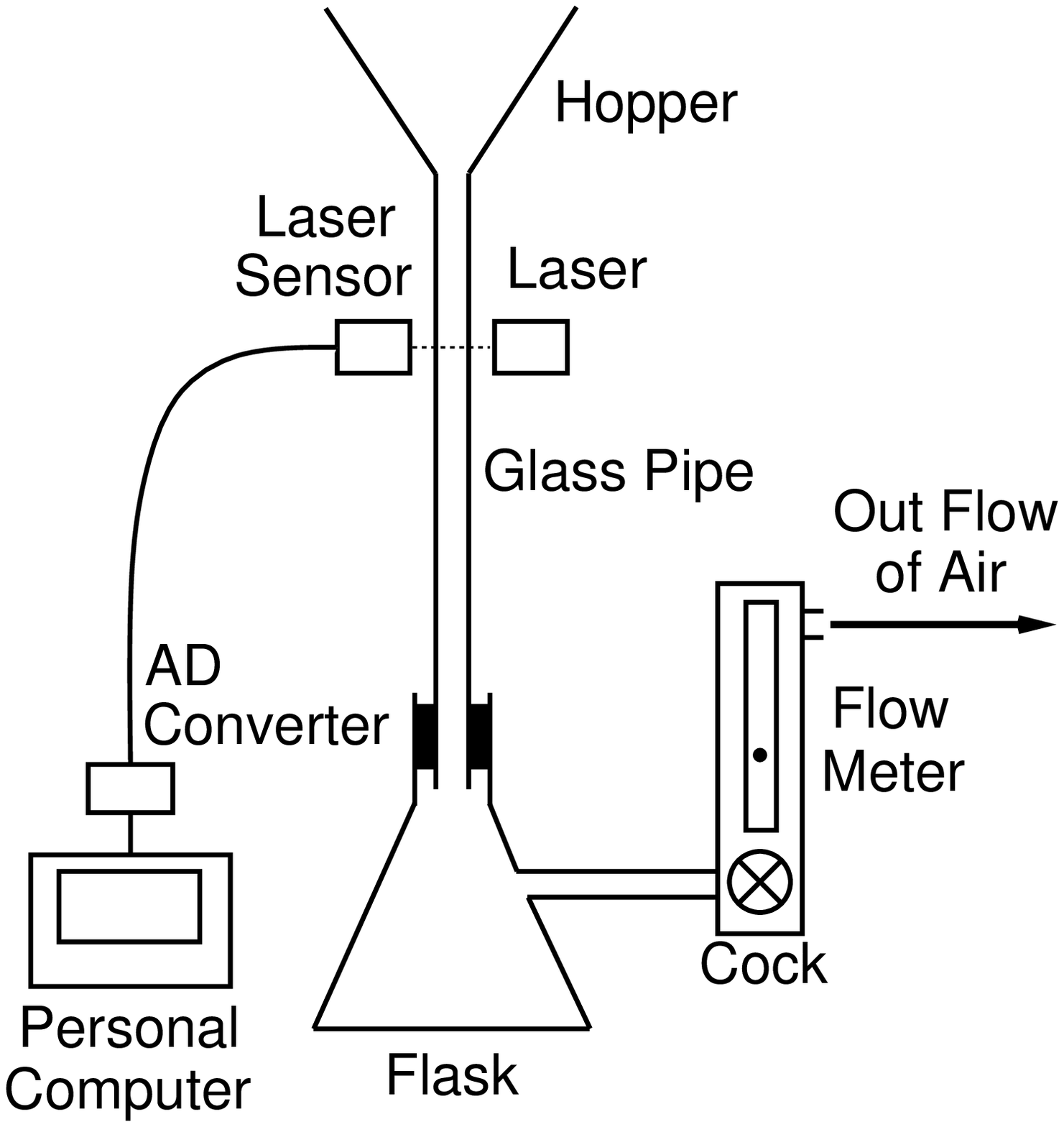,width=6cm,height=5cm}
\caption{Schematic illustration of the experimental setup.}
\end{center}
\end{figure}

  Our experimental setup is shown in Fig. 1.  We used a glass pipe of 
1500 mm length and 3 mm inner diameter, {\it i.e.}, the same one used in 
Ref. \cite{horikawa95,horikawa96}.  A flask is connected to the bottom end of 
the vertical pipe and a flow-meter is attached to an outlet stuck out of 
the flask.  We used rough sand, {\it i.e.}, our granular particles are 
polydisperse and their average diameter is about 0.3 mm.  Thus the inner 
diameter of the pipe is about ten times larger than the average diameter of 
sand.  We pored rough sand into a hopper connected to the top end of the pipe, 
which flows through the pipe 
due to gravity.  
Sand finally falls into the flask while air can exit 
out of it through the flow-meter, which can control the rate of air 
discharge.  Meanwhile we measured 
density fluctuations as the transmission light intensity across the pipe 
at a fixed location midway up the pipe
by using laser light and detecting system (KEYENCE, LX2--02).  The laser 
light has a rectangular 
cross-section of 10 mm wide and 1 mm high and is emitted with a pulse 
frequency of 4096 Hz.  The detecting system has linear response between the 
output voltage and the density of particles in the range of 
output voltage between 1 and 5 volts.

Let us now consider what happens by closing the cock of flow-meter.  
Granular particles falling in a pipe interact with the medium (air) due 
to the viscosity.  When the flow-meter is removed from the apparatus, 
{\it i.e.}, the bottom end of the pipe is fully open, granules can fall rather 
freely as if the existence of air could be neglected.  (Hereafter we will 
refer this situation to {\it fully open}.)  In {\it fully open} 
there are no visible density waves, as reported in Refs. 
\cite{horikawa95,horikawa96}.  The reason 
is that both granules and air flow together through the pipe.  As the 
cock is gradually closed, however, the pressure in the flask rises and the 
effect of the viscous force becomes more important.  In particular, when the 
cock is fully closed (we refer to {\it fully closed}), air in the 
pipe must go upward due to approximate conservation of the total volume 
(sand plus air) in the flask while sand falls downward.  We consider 
that the increment of interaction 
between granules and air induces density waves.

\begin{figure}[htbp]
\begin{center}
\epsfile{file=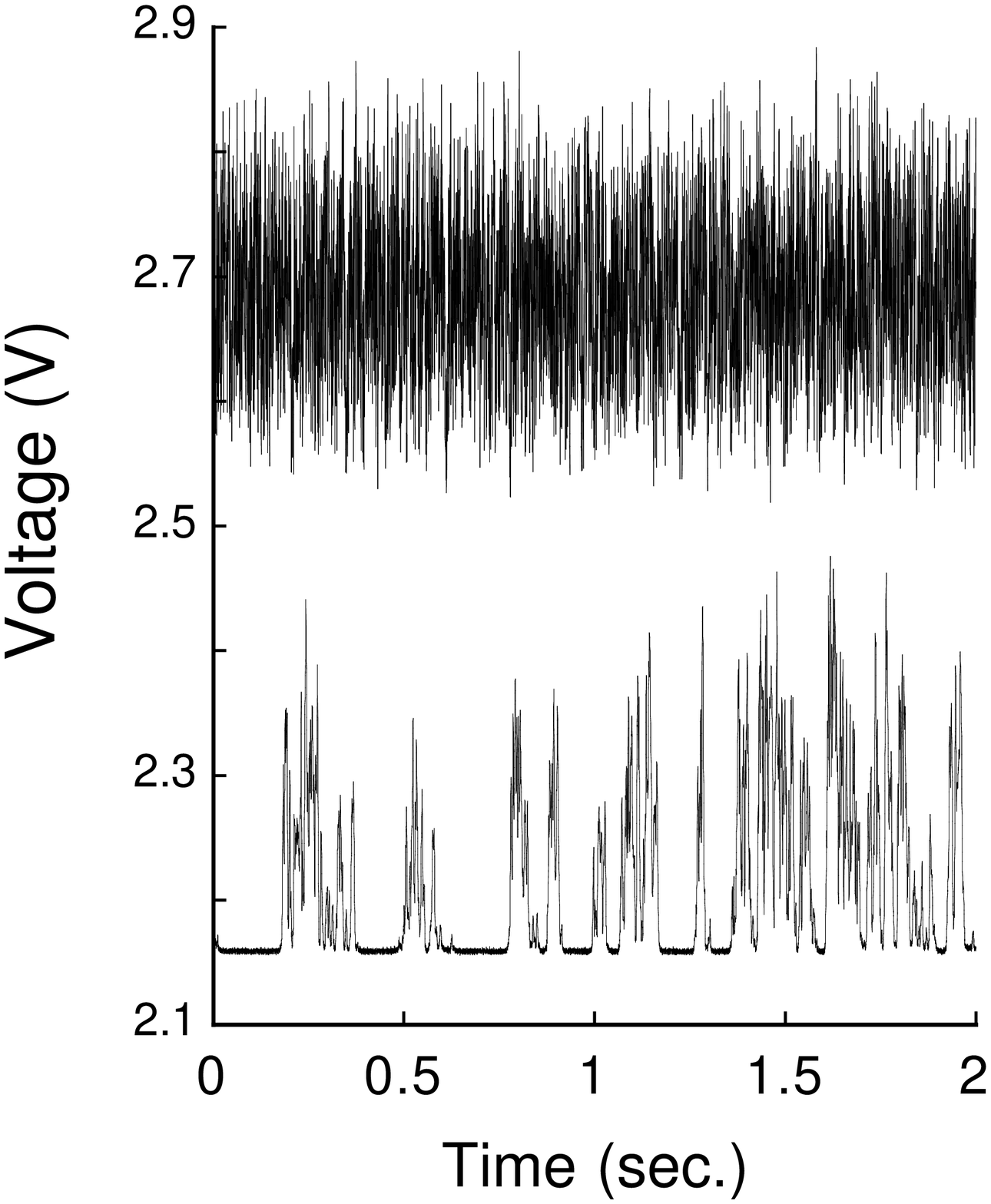,width=7cm,height=7cm}
\caption{Time series signal of the granular flows for {\it fully open} (top) 
and {\it fully closed} (bottom).  Upper signal is shifted by 0.4 V to avoid 
the data overlap}
\end{center}
\end{figure}


Figure 2 shows typical time series signals of granular flows 
measured at $x = 100$ cm, where $x$ is defined as the distance of 
measuring point along the pipe from the top.  In contrast with the 
white-noise-like signal (top) in {\it fully open}, one can see that the 
density wave for {\it fully closed} (bottom) has intermittent structure.  
In this 
figure higher (lower) voltage corresponds to smaller (larger) granular density 
since we measured the light intensity transmitted 
across the pipe.  In particular, the output voltage is about 2.15 volts when 
a pipe is fully packed with sand, and about 2.50 volts when it is completely 
empty.  This assures us that our measurements were well in the linear 
range between the voltage and the density.  In the case of real flow patterns  
one can observe that two clusters sometimes collide 
with each other and merge into one, and sometimes one cluster splits into more 
than one cluster, in a behavior reminiscent of a chain of traffic jams on a 
crowded highway\cite{bando,kerner93,komatsu95}.  In the following 
discussion of this Letter we restrict ourselves to  {\it fully closed}.  
(When the medium air was controlled to flow downward by 
gradually opening the cock of the flow-meter, 
the position of the emergence of clear density waves gradually 
shifted downward.  The details of the results will be presented elsewhere.)

\begin{figure}[htbp]
\begin{center}
\epsfile{file=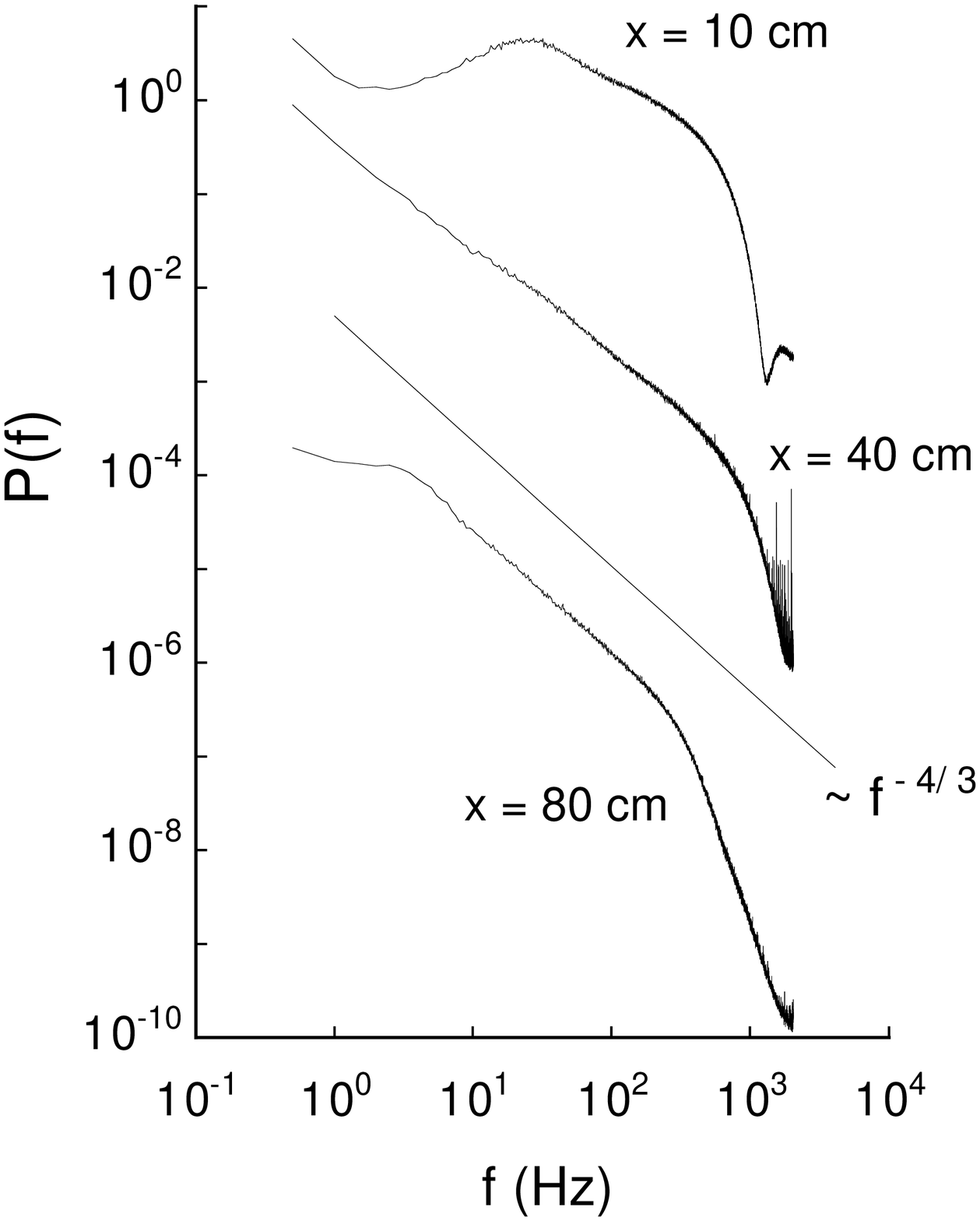,width=8cm,height=9cm}
\caption{Log-log plot of power spectra $P(f)$ of time series signals 
in {\it fully closed}.  The straight line with the slope of $-4/3$ in the 
figure is for an eye-guide.}
\end{center}
\end{figure}

  Let us pay our attention to the power spectrum $P(f)$ of density 
fluctuations of granular flows.  
Figure 3 shows $P(f)$ of recorded signals in 
{\it fully closed}.  Each spectrum was obtained by averaging over 640 
independent data with length 8192 discrete points each (two seconds in real 
time) and was appropriately shifted to avoid the overlap.  
The self-organization of the power-law form $P(f) \sim f^{-\alpha}$ is 
observed as the measuring position $x$ increases: The spectrum at $x = 10$ cm 
is similar to that in {\it fully open}.  Density waves have already 
emerged at $x = 40$ cm where the scaling regime is covered by 
the whole frequency range except for the fast decay in higher $f$.  Then 
the system falls into the steady state from about $x = 50$ cm downward.  The 
scaling 
range for the steady state, {\it i.e.}, for $x \geq 50$ cm is from 10 Hz to 
200 Hz, as seen in Fig. 3.

In the frequency range from 15 Hz to 150 Hz, values of the slope are fitted by 
the least mean square method as $\alpha = 1.33 \pm 0.04, 1.31 \pm 0.05$ and 
1.34 $\pm$ 0.05 for $x = 80$, 90 and 100 cm, respectively.  (The data at 
$x = $90 cm and 100 cm are not shown in Fig. 3.)  These values are 
very close to 4/3 suggested by 
LGA simulation\cite{peng94}.  
%
%

  Let us allocate the rest of this Letter to 
introduce a 
one-dimensional model which reproduces $\alpha = 4/3$
in {\it fully closed} system.

  The important mechanisms for particle dynamics are 
the drag between air and particles, and
the relaxation process to an optimal velocity
which may be the sedimentation rate of particles.
  When the $N$ particles are confined in a quasi-one dimensional container, 
the motion of particles may be described by the following nonlinear equation:
\begin{equation} \label{1d_powder}
\ddot r_{n} + \zeta [\dot r_{n} - W(\{r_n\})]=T F(\{r_n\})
 + f_{n}(t),
\end{equation}
where $r_{n}$ and $\zeta$ are the relative distance between the $n$-th 
and $(n+1)$-th particles, 
and the drag coefficient,  respectively.  
The collisional force
$F(\{r_n\})=\varphi'(r_{n+1})+\varphi'(r_{n-1})-2\varphi'(r_{n})$ comes from
a soft core repulsive potential $\varphi(r_n)$.
The parameter $T$ represents the strength of repulsion.
  The optimal velocity 
$W(\{r_n\})=U(\frac{r_{n+1}+r_n}{2})-U(\frac{r_n+r_{n-1}}{2})$
   is the linear combination of sedimentation rate
$U(x)$ which is  the nonlinear function of the local 
volume fraction\cite{sedi} in general. 
  The most crucial simplification of  (\ref{1d_powder})  without
any justification
is to assume that $f_{n}$ is a Gaussian white noise with zero mean.
It should be noticed that 
the  drag $\zeta$ is irrelevant in {\it fully open},
because air in the pipe flows away together with particles.
Thus, the motion of particles are almost elastic and
clear density waves cannot be observed in {\it fully open}. 
It should be noted that (\ref{1d_powder}) is written for the 
relative motion of particles. 
Discarding the noise term from (\ref{1d_powder}), we obtain
an equation of motion for particle at the position $x_n$ as
$\ddot x_n +\zeta [\dot x_n-U(\frac{x_{n+1}-x_{n-1}}{2})]=
T[\varphi'(r_n)-\varphi'(r_{n-1})]$
where  $r_n=x_{n+1}-x_n$.

Linearizing (\ref{1d_powder}) around the uniform solution $\dot r_n=0$ where 
$a= \bar r_n\equiv N^{-1}\sum_n r_n$, we obtain 
\begin{equation}\label{linear}
\ddot{\tilde r_k} +\zeta[\dot{\tilde r_k} 
-i  U'\sin k \tilde r_k]=2T\varphi''
(\cos k-1)\tilde r_k+\tilde f_k(t),
\end{equation}
where the argument of $U'$ and $\varphi''$ is $a$.
$\tilde r_k$ and $\tilde f_k(t)$ are
respectively  the Fourier transform of  $\delta r_n=r_n-a$
and $f_n(t)$.
Equation (\ref{linear}) has the solution $\tilde r_k(t)\propto \exp[\sigma_{\pm}t]$
, where 
\begin{equation}\label{sigma}
\sigma_{\pm}=-\frac{\zeta}{2}\pm \displaystyle\sqrt{\left(\frac{\zeta}{2}
\right)^2-2T\varphi''(1-\cos k)+i\zeta  U'\sin k}.
\end{equation}
${\rm Re}[\sigma_+]$ represents the relevant eigenvalue of the linear problem,
which becomes positive for 
$ U'(a)^2\cos^2(k/2)\ge T \varphi''(a)$.  Thus the most 
unstable wave number is $k\to 0$ and  the neutral curve is given by 
$T_c= U'^{2}/\varphi''(a)$. 
At $T=T_c(1-\mu)$ the expansion of $\sigma_+$ around $k=0$
is given by 
$\sigma_+(k)\simeq i[ c_0 k- \displaystyle\frac{c_0}{6} k^3+\cdots]
          + \displaystyle\frac{c_0\mu}{ \zeta} k^2
- \displaystyle\frac{c_0^2}{4\zeta}k^4 +\cdots
$, 
where $c_0= U'(a)$. 
Thus, for $\mu>0$ the uniform state is
unstable due to  the negative diffusion.

Adopting  
$U(r)=\tanh(r-2)+\tanh(2)$, $\varphi(r)={\rm sech}^2(r)$
, $\zeta=2$,  $N=256$,
$T_c=3.95798\cdots$,
and $a=2$ at $t=0$,
we simulated (\ref{1d_powder}) 
by the classical Runge-Kutta method 
 until $t=2^{11}$ with time interval $\Delta t=1/2^4$
under the periodic boundary condition. 
We used 
 the uniform random number distributed between $- X$ and $X$ 
with $X=9/1024$\cite{value} for $f_n(t)$.
 Figure 4 displays  
the power spectrum $P(f)=|\tilde c(f)|^2$ 
obtained from our simulation of (\ref{1d_powder}) at $\mu=1/64$, 
where $\tilde c(f)$ is the Fourier
transform of  the discretely sampled data of the density 
$c(t)=\displaystyle\frac{1}{N}\sum_{n} \frac{1}{r_n(t)}$ with the interval 1.
This clearly supports $P(f)\sim f^{-4/3}$ as in 
our experiment.  
From the examinations of several  values of $\mu$, 
we have confirmed that the qualitative results
are insensitive to  the sign of $\mu$ when
$|\mu|\ll 1$.  This result is reasonable because near the neutral curve
the time scale of relaxation or growth of fluctuations is much longer than
the  time scale induced by the noise $f_n(t)$.
Our result suggests that the linear relaxation 
theory of fluctuations 
 can be used to explain $P(f)\sim f^{-4/3}$.

\begin{figure}[htbp]
\begin{center}
\epsfile{file=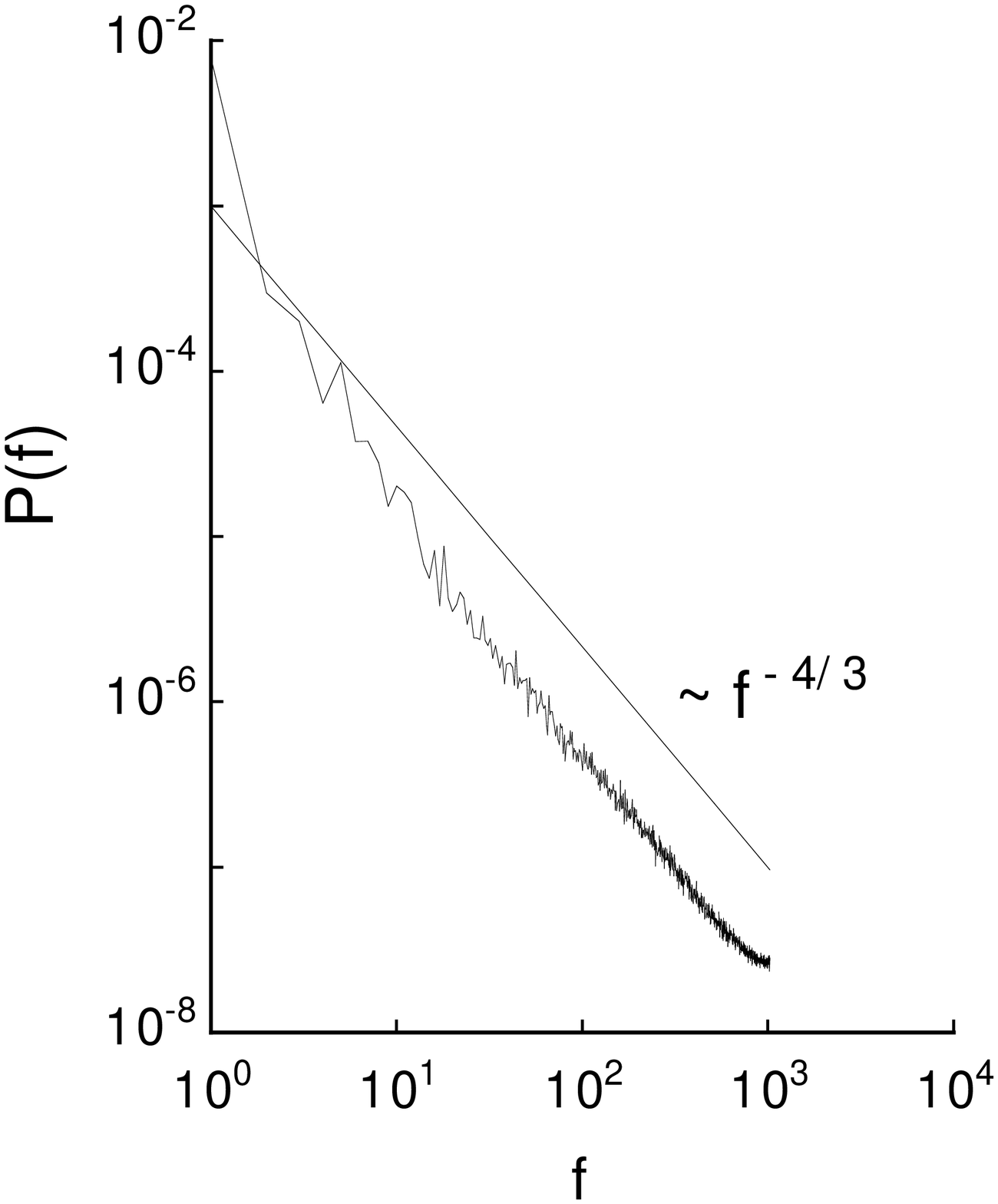,width=7cm,height=7cm}
\caption{Log-log plot of power spectrum $P(f)$ obtained from
the numerical integration of (1), where the unit of $f$ is $1/(2\pi)$. 
The guide line represents $f^{-4/3}$. See the text
for details.}
\end{center}
\end{figure}
 
Let us briefly sketch how to derive 4/3-law from
the  behavior of structure factor   
\begin{equation}\label{scattering}
S_k(t)=
\sum_{n,m}<\exp[i k(\delta  r_n(t)-\delta  r_m(0))]> 
\end{equation}
in weakly stable states, 
{\it i.e.}, $\mu<0$ and $|\mu|\ll 1$,
where $\delta r_n=r_n-a$ is the fluctuation of relative distance.
The structure factor can be rewritten as 
$S_k(t)=
\frac{1}{N}\sum_{n, m}\exp \left[-\frac{k^{2}}{2}\phi_{nm}(t)\right]$
where
$\phi_{nm}(t)=<(\delta r_n(t)-\delta r_m(0))^2>$.
For $\mu<0$,  $S_k(t)$ can be  
calculated
 as in the case of polymer dynamics\cite{com}.
With the aid of the expansion of $\sigma_+$, (\ref{1d_powder}) is
reduced to
\begin{equation}\label{4}
\partial_{\tau}r(z,\tau)-\partial_z^3 r(z,\tau)=\epsilon 
[\partial_z^2-\partial_z^4]
r(z,\tau)+\xi(z,\tau)
\end{equation}
where $\tau=\epsilon^3\beta t$, 
$z=\displaystyle\frac{2\zeta}{3c_0}\epsilon (x+c_0t)$,
$\xi(z,\tau)=\epsilon^3 \beta f_n(t)$ with 
$\epsilon=\displaystyle\frac{3\sqrt{c_0}}{\zeta}\sqrt{-\mu}$ and
 $\beta=\displaystyle\frac{4}{3\sqrt{c_0}}$.  
The solution of  (\ref{4}) 
is given by
$ \tilde r_k(\tau)\simeq \int_{0}^{\tau}ds
\exp[\lambda_k(\tau-s)]\tilde \xi_k(s),
$
where $\lambda_{k}=i k^3 -\epsilon k^2(1+k^2)$.
Thus, we obtain the correlation
\begin{equation}\label{r-r}
< \tilde r_k(\tau) \tilde r_{-k}(0)>
=\frac{D}{2\epsilon l k^2(1+k^2)}
\exp[\lambda_k\tau],
\end{equation}
where $l$ is the system size, and 
we used $<\tilde \xi_k(\tau)\tilde \xi_p(\tau')>=
\frac{D}{l}\delta_{k+p,0}\delta(\tau-\tau')$.

A long but straightforward calculation parallel to Ref.\cite{com}
yields
\begin{equation}\label{form}
S_k(\tau)\simeq 
\int_{-l}^l dx \exp[-D_G k^2\tau-\frac{D k^2}{2\epsilon}w -
\frac{D k^2}{\pi \epsilon}\tau^{1/3} h(u)], 
\end{equation}
where the argument of $S_k$ is replaced by the scaled time, 
$D_G$ is the diffusion constant for the gravitational center
in (\ref{4}), $u=x\tau^{-1/3}$,
and $w = |z-z'|$.
Since  $h(u)$  converges to $h(0)=\pi/\Gamma(1/3)$ 
 as time goes on,  
we obtain
\begin{equation}\label{result}
S_k(\tau)\simeq \frac{\epsilon}{D k^2}
\exp[-\frac{D k^2}{\epsilon\Gamma(1/3)}\tau^{1/3}]
\end{equation}
in intermediate time range.
In the limit of small $\tau$, 
 $S_k(\tau)\propto 1- 
\displaystyle\frac{4D k^2}{\epsilon\Gamma(1/3)}k^2\tau^{1/3}+\cdots$. 
Thus its 
Fourier transform, giving the power spectrum $P_k(f)$
obeys
\begin{equation}\label{spectrum}
P_k(f)\sim f^{-\alpha}, \quad \alpha = 4/3 \quad 
({\rm as}\quad f \to \infty),
\end{equation}
where use was made of 
$\int_{-\infty}^{\infty} dt e^{i 2\pi f t}|t|^{1/3}\propto f^{-4/3}$.
The value $4/3$ is identical to the one obtained by our experiments and 
numerical simulations\cite{peng94}.  Thus our model 
(\ref{1d_powder}) reproduces $\alpha = 4/3$.
It should be noted that the appearance 
of this power-law form in the original model (\ref{1d_powder}) is only for 
$f < \zeta$ since we eliminate the fast decaying mode $\sigma_{-}$ in our 
analysis.  This tendency is also observed as the higher-frequency cutoff 
in our experiment (see Fig. 3).

The 4/3-law is determined by short time behavior of 
the  dynamics of density waves induced by the noise,
which  
is essentially determined by the 
linear dispersion relation $\lambda_k\sim i k^3$.  
The details of our theoretical analysis including 
the effects of nonlinearity
 will be discussed elsewhere\cite{nisi}.

  In this Letter we have confirmed  $\alpha = 4/3$ as 
the power-law exponent in the frequency spectrum of density correlation 
function from  the experiment, the simulation and the theory. 
We have also clarified the mechanism to yield $f^{-4/3}$ spectrum
which is related to the critical slowing down of the density fluctuations.
 It should be noticed that the continuous increase of
$\alpha$ in LGA\cite{peng95} from $\alpha=0$ to 2
with the particle density is consistent with the 4/3-law and our picture,
 because the spectrum determined by the noise
in linearly stable uniform state far from the neutral curve
should be white ($\alpha=0$) and
 the effective exponent of the power-law
becomes large when the exponential decay ({\it i.e.} $\alpha=2$)
in the off-critical region exists.
There is, however, a 
discrepancy between our results and the one on experiments in liquids
\cite{nakahara97}.  The reason of this difference should be clarified in
the future.

  OM, NK and MM would like to thank Y-h. Taguchi, T. Isoda and A. Nakahara 
for stimulating discussions.  HH thanks K. Ichiki, K. Nakanishi 
and G. Peng for 
fruitful discussions.  
This work is partially supported by Grant-in-Aid for Science Research Fund 
from the Ministry of Education, Science and Culture (09740314).





\begin{references}
\bibitem{jaeger96}
H. M. Jaeger, S. R. Nagel and R. P. Behringer, Rev. Mod.\  
Phys.\ {\bf68}, 1259 (1996).
\bibitem{taguchi}Y-h. Taguchi, Phys. Rev. Lett. {\bf 69}, 1367 (1992), 
K. M. Aoki T. Akiyama, Y. Maki and T. Watanabe, Phys. Rev. E {\bf 54}, 874 
(1996).
\bibitem{rosato87}A. Rosato, K. J. Strandburg, F. Prinz and R. H. Swendsen, 
Phys.\ Rev. Lett. {\bf 58}, 1038 (1987).
\bibitem{pak94}H. K. Pak and R. P. Behringer, Nature {\bf 371}, 231 (1994). 
\bibitem{swinney95}F. Melo, P. B. Umbanhowar and H. L. Swinney, Phys.\ Rev. 
Lett. {\bf 75}, 3838 (1995).
\bibitem{swinney96}P. B. Umbanhowar, F. Melo and H. L. Swinney, Nature 
{\bf 382}, 793 (1996).
\bibitem{batchelor88}G. K. Batchelor, J. Fluid Mech. {\bf 193}, 75 (1988).
\bibitem{sasa92}S. Sasa and H. Hayakawa, Europhys. Lett.\ {\bf 17}, 685 
(1992), T. S. Komatsu and H. Hayakawa, Phys. Lett. A {\bf 183}, 56 (1993).
\bibitem{peng94}G. Peng and H. J. Herrmann, Phys. Rev. E {\bf 49}, R1796 
(1994).
\bibitem{peng95}G. Peng and H. J. Herrmann, Phys. Rev. E {\bf 51}, 1745 (1995).
\bibitem{horikawa95}S. Horikawa, A. Nakahara, T. Nakayama and M. Matsushita, 
J. Phys. Soc. Japan {\bf 64} 1870 (1995).
\bibitem{horikawa96}S. Horikawa, T. Isoda, T. Nakayama, A. Nakahara and 
M. Matsushita, 
Physica A {\bf 233}, 699 (1996).
\bibitem{nakahara97}A. Nakahara and T. Isoda, Phys. Rev. E {\bf 55}, 4264 
(1997).
\bibitem{bando}M. Bando, K. Hasebe, A. Nakayama, A. Shibata and Y. Sugiyama, 
Phys. Rev. E {\bf 51}, 1035 (1995).
\bibitem{kerner93}B. S. Kerner and P. Konhauser, Phys. Rev. E {\bf 48}, 2335 
(1993).
\bibitem{komatsu95}T. S. Komatsu and S. Sasa, Phys. Rev. E {\bf 52}, 5574 
(1995).
\bibitem{sedi} H.Hayakawa and K.Ichiki, Phys.Rev. E {\bf 51}, R3815 (1995)
and references therein.
\bibitem{value} The value of $X$ is chosen to have the unit amplitude
in the form of eq.(5).
\bibitem{com} P.G. de Gennes, Physics {\bf 3}, 37 (1967): see also
M. Doi and S. F. Edwards, {\it The Theory of Polymer Dynamics} 
(Oxford, 1986).
\bibitem{nisi} Detailed calculation  within the linear theory
discussed here can be seen in :
H. Hayakawa and K. Nakanishi, Prog.Theor.Phys. Suppl. (to be published).

\end{references}
\end{document}